\documentclass[pre,twocolumn,aps,longbibliography]{revtex4-2}

\usepackage{amssymb}
\usepackage{tabularx}
\usepackage{multirow}
\usepackage[colorlinks]{hyperref}
\usepackage{soul,color}
\usepackage{graphicx}
\usepackage{dcolumn}
\usepackage{bm}
\usepackage{subfigure}

\usepackage{footmisc}

\newcommand{\appropto}{\mathrel{\vcenter{
  \offinterlineskip\halign{\hfil$##$\cr
    \propto\cr\noalign{\kern2pt}\sim\cr\noalign{\kern-2pt}}}}}

\begin{document}


\title{Memristive Ising Circuits}


\author{Vincent~J.~Dowling}
\affiliation{Department of Physics and Astronomy, University of South Carolina, Columbia, SC 29208 USA}

\author{Yuriy~V.~Pershin}
\email{pershin@physics.sc.edu}
\affiliation{Department of Physics and Astronomy, University of South Carolina, Columbia, SC 29208 USA}



\begin{abstract}
The Ising model is of prime importance in the field of statistical mechanics. Here we show that Ising-type interactions can be realized in periodically-driven circuits of stochastic binary resistors with memory. A key feature of our realization is the simultaneous co-existence of ferromagnetic and antiferromagnetic interactions between two neighboring spins -- an extraordinary property not available in nature. We demonstrate that the statistics of circuit states may
perfectly match the ones found in the Ising model with ferromagnetic or antiferromagnetic interactions, and, importantly, the corresponding Ising model parameters can be extracted from the probabilities of circuit states. Using this finding,
the Ising Hamiltonian is re-constructed in several model cases, and it is shown that different types of interaction can be realized in circuits of stochastic memristors.
\end{abstract}


\maketitle

\section{Introduction}

The utilization of electronic circuits as an analog to other physical systems is becoming more and more prevalent. It has been recently shown that certain circuits comprised of only capacitors and inductors~\cite{LeeTopolectric18,Imhof18} as well as circuits combining passive resistive~\cite{Hofmann20,DiVentra22a} or active~\cite{Kotwal21a} components with capacitors and inductors can be used to realize the same states that are found in topological phases in condensed matter~\cite{Asboth2016,ShenTI,Tukoup20,Hadad16}, forming a connection between two, otherwise, distinct systems. For instance, in the topoelectric Su-Schrieffer–Heeger (SSH) circuit~\cite{LeeTopolectric18} the boundary resonances  in the impedance are reminiscent of edge states in the SSH model. Here, we introduce a circuit of stochastic memristors exhibiting the same statistics of states as in the Ising model.

\begin{figure}[b]
(a) \includegraphics[width=0.42\columnwidth]{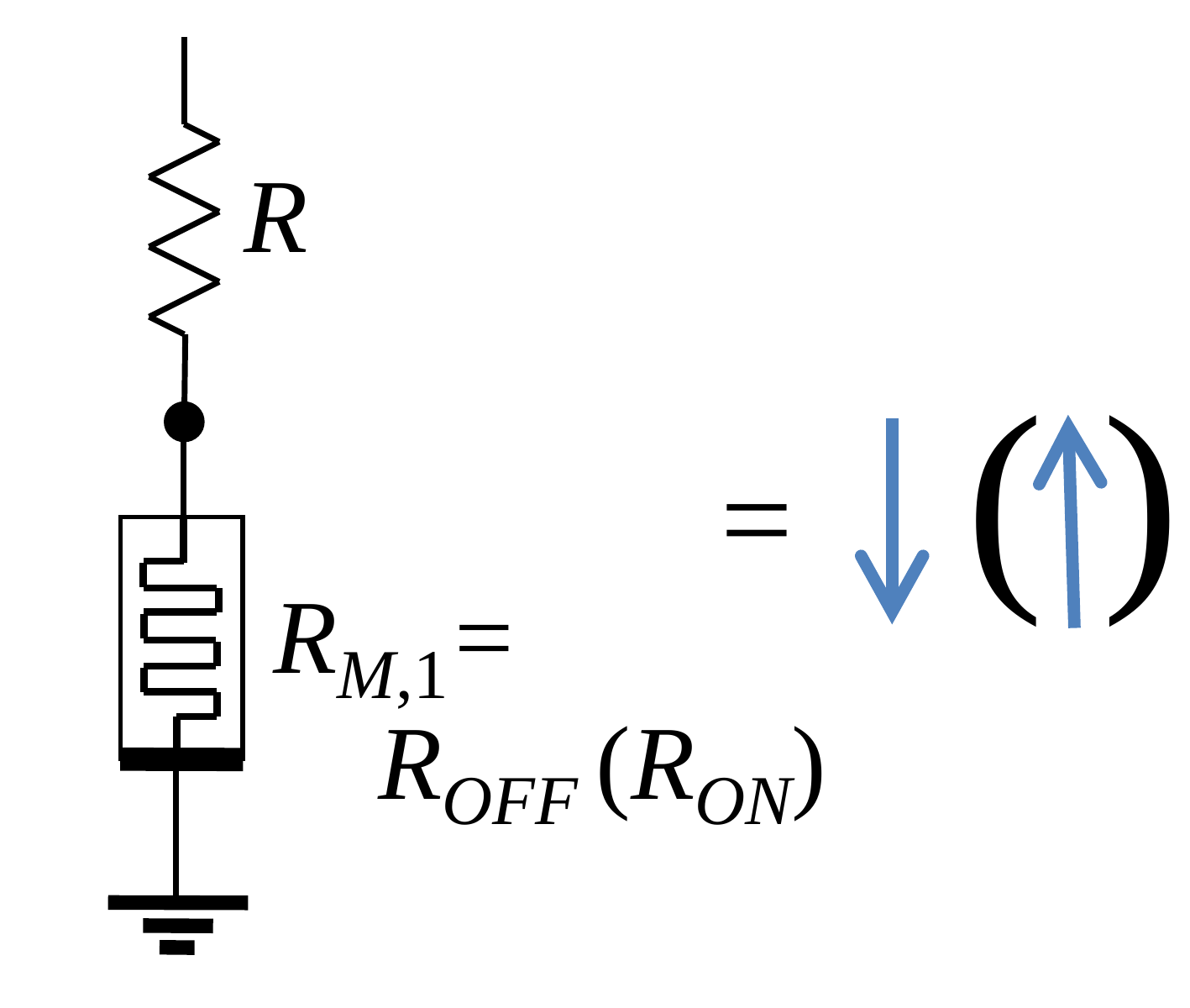} \hspace{1mm} (c) \includegraphics[width=0.42\columnwidth]{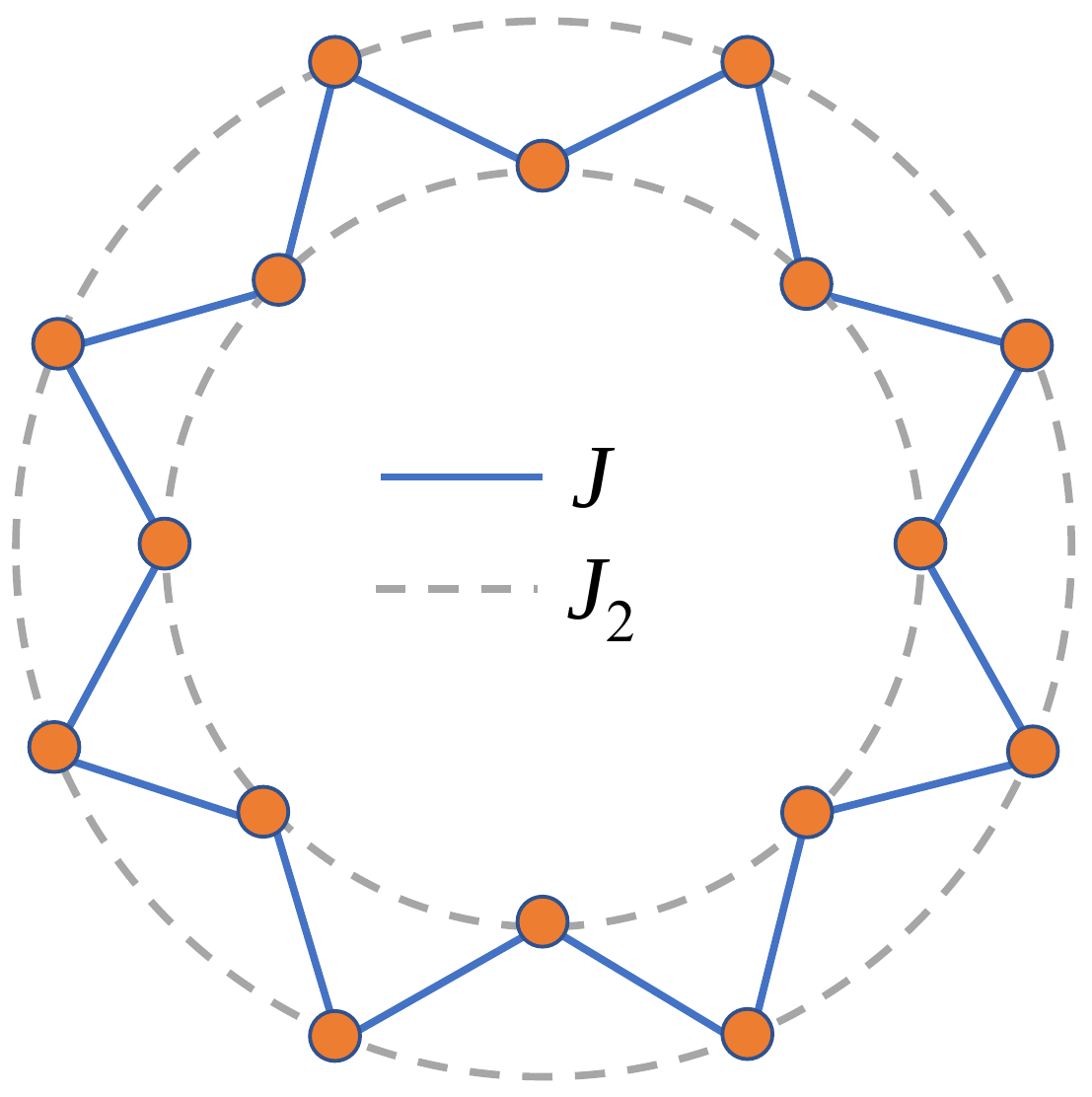} \\
\vspace{0.6cm}
(b)\hspace{2mm} \includegraphics[width=0.9\columnwidth]{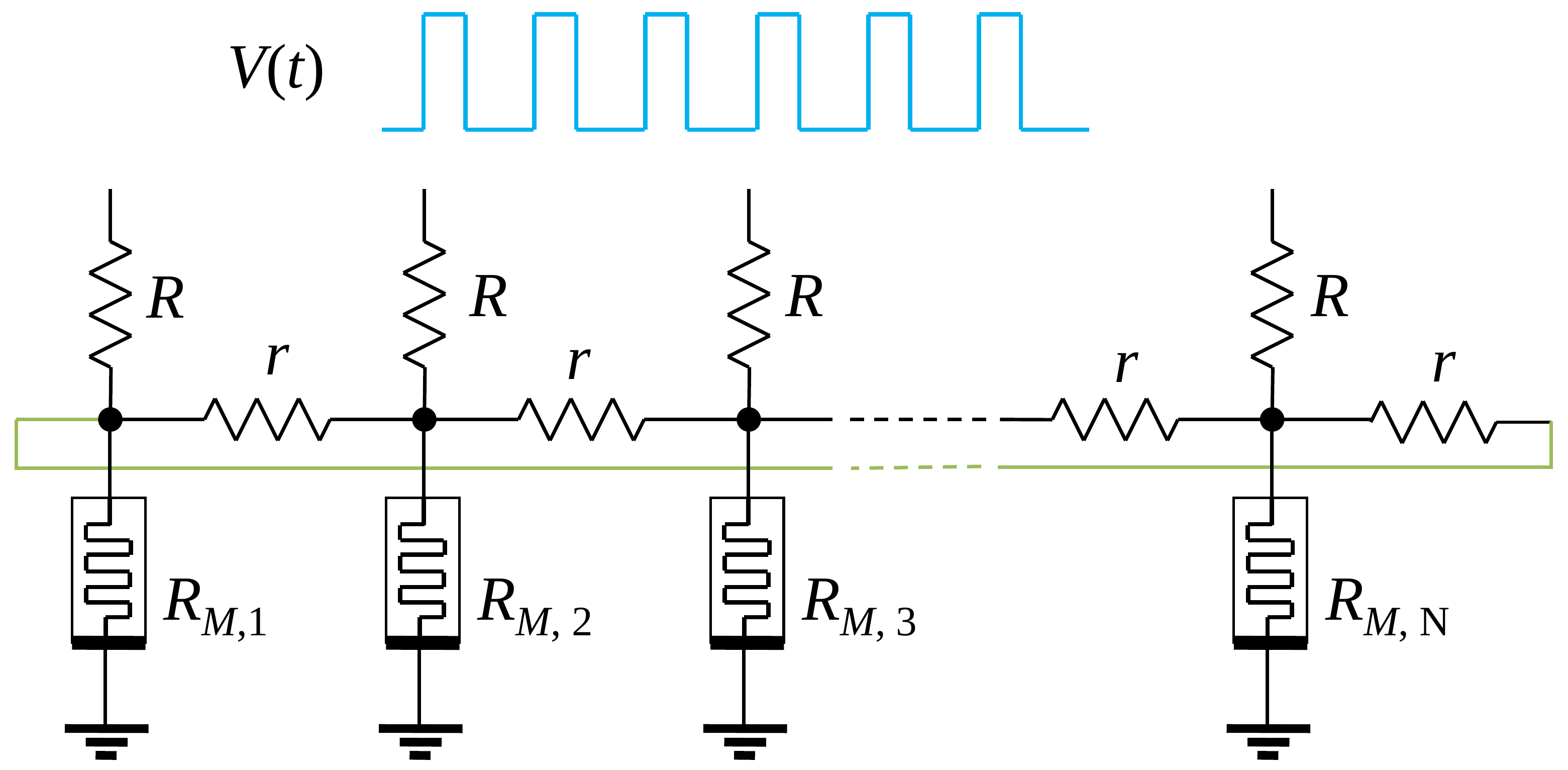}
\caption{(a) Memristive spin sub-circuit: the high- and low-resistance states of stochastic memristor correspond to spin-down (0) and spin-up (1) states, respectively. (b) One-dimensional memristive Ising circuit with a periodic boundary condition. Here, $r$-s denote the resistance of coupling resistors. (c) The scheme of interactions in the Ising Hamiltonian. \label{fig:1}}
\end{figure}

While the concept of constructing an electric analog to the Ising model is not novel~\cite{Mahboob16,aadit2022,borders2019,sutton2017,chou2019,dutta2019,dutta2021,cai2020,bojnordi2016} and gaining increasing attention in the context of building Ising machines~\cite{aadit2022,borders2019,sutton2017,chou2019,dutta2019,dutta2021,cai2020,bojnordi2016}, our approach is. The basic idea is as follows. We use a resistor and stochastic memristor connected in-series as a memristive spin (Fig.~\ref{fig:1}(a)), and couple memristive spins by resistors to induce their interactions (see Fig.~\ref{fig:1}(b) for the circuit considered in this Letter).
It is assumed that the stochastic memristor can be found in one of two states, $R_{ON}$ and $R_{OFF}$ (such that $R_{ON}<R_{OFF}$), and the switching between these states occurs probabilistically and is described by voltage-dependent switching rates (the details of the model are given below). The circuit is subjected to alternating polarity pulses that drive the memristive dynamics.  The states of memristors are read each period of the pulse sequence (say, at the end of the negative pulse) and the probabilities of these states are determined. We note that the circuit in Fig.~\ref{fig:1}(b) but with deterministic memristors was introduced in Ref.~\cite{Slipko17a}, and a mean-field model of memristive interactions in a similar (but not the same) deterministic circuit was developed in~\cite{caravelli2018}.

 Using numerical simulations, we have found that our circuit is capable of exhibiting an analogous type of ordering in memristor configurations as of those found in magnetic materials. Meaning, there can exist a strong bias for a specific circuit to exist in an antiferromagnetic (AFM) memristor configuration ($-R_{ON}-R_{OFF}-R_{ON}-R_{OFF}-$) or an ferromagnetic (FM) memristor configuration ($-R_{ON}-R_{ON}-R_{ON}-R_{ON}-$ or $-R_{OFF}-R_{OFF}- R_{OFF}-R_{OFF}-$). In fact, a very important aspect of our circuit is the simultaneous co-existence of AFM and FM interactions between two neighboring spins. The goal of this work is to demonstrate the possibility of the standard magnetic orderings (AFM and FM) in the memristive Ising circuits.

This paper is organized as follows. In Sec.~\ref{sec:2} we introduce the Ising Hamiltonian, stochastic model of memrisotrs, and make the connection between the statistical properties of the circuit and ones of the Ising Hamiltonian. In the same section, we briefly discuss the numerical approach used in our work. The results of our simulations are presented in Sec.~\ref{sec:2} with the emphasise on the possibility of reaching FM and AFM interactions in the circuit. The paper ends with a conclusion.

\section{Methods} \label{sec:2}

Mathematically, we utilize an effective Ising-type Hamiltonian to describe the probabilities observed in the circuit simulations.  For the circuit in Fig.~\ref{fig:1}(b), the Hamiltonian has the form
\begin{equation}\label{eq:1}
  H=-J\sum_{i}\sigma_i\sigma_{i+1}-J_2\sum_{i}\sigma_i\sigma_{i+2}-h\sum_i\sigma_i\;\;,
\end{equation}
where $J$ is the interaction coefficient for adjacent spins, $J_2$ is the next-to-adjacent interaction, $h$ is the magnetic field, and periodic coupling is assumed. Schematically, these interactions are presented in Fig.~\ref{fig:1}(c).  We consider the electronic circuit as a physical system described by the Boltzmann distribution
\begin{equation}\label{eq:2}
  p_i=\frac{1}{Z}e^{-\frac{E_i}{kT}}.
\end{equation}
Here, $Z=\sum\limits_{j}e^{-\frac{E_j}{kT}}$ is the statistical sum, and $E_j$-s are the ``energies'' of circuit states. We argue that for the circuit in Fig.~\ref{fig:1} and similar circuits these ``energies'' correspond to the Ising Hamiltonian~(\ref{eq:1}).

 To explain the co-existence of AFM and FM interactions, consider a set of identical memristors in $R_{OFF}$ subjected to a positive voltage pulse driving the OFF-to-ON transition. Each memristor will have an equivalent probability of being the first to switch states. When one of these memristors swaps states, it
 reduces the probability to switch of its neighbors (reducing the voltage across them). In this scenario, memristors with neighbors both in the $R_{OFF}$ state will have the highest chance of switching.  This leads to the tendency of antiferromagnetic ordering in the memristors under a positive voltage pulse. However, under a negative voltage pulse the $R_{ON}$ state memristors with neighboring $R_{OFF}$ state memristors will be favored to change states. Meaning, the configuration will tend towards ferromagnetic ordering under a negative voltage pulse. The overall ordering of a memristive circuit driven by an AC source will then be dependent on the choice of model parameters for the memristors. Based on the parameters, one type of ordering may be dominant.



Next, we introduce the model of stochastic memristors. According to experiments with certain electrochemical metallization (ECM) cells~\cite{jo2009programmable,gaba2013stochastic} and valence change memory (VCM) cells~\cite{naous2021theory}, the probability of switching between resistance states of these devices can be described by switching rates of the form
\begin{eqnarray}
 \gamma_{0\rightarrow 1}(V) &=& \left\{ \begin{array}{ccl}
\left( \tau_{01} e^{-V/V_{01}}\right)^{-1}&,& V>0   \\
0 &,& \textnormal{otherwise}
\end{array}\right. \;\; , \label{eq:gamma01} \\
 \gamma_{1\rightarrow 0}(V) &=& \left\{ \begin{array}{ccl}
\left( \tau_{10} e^{-|V|/V_{10}}\right)^{-1}&,& V<0 \\
0 &,& \textnormal{otherwise}
\end{array}\right. \; , \label{eq:gamma10}
\end{eqnarray}
where $V$ is the voltage across the device, and $\tau_{01(10)}$ and $V_{01(10)}$ are device-specific parameters. Here, 0 and 1 correspond to the high ($R_{OFF}$) and low ($R_{ON}$) resistance states, respectively. Under a constant voltage, the probability to switch follows the distribution~\cite{jo2009programmable,gaba2013stochastic}
\begin{equation}
P(t)=\frac{\Delta t}{\tau(V)}e^{-t/\tau(V)}, \label{eq:distrib}
\end{equation}
where $\tau(V)$ is the inverse of the switching rate given by Eq.~(\ref{eq:gamma01}) or (\ref{eq:gamma10}) (depending on the sign of $V$). Previously, we have developed a master equation approach for the circuit of stochastic memristors~\cite{dowling2020probabilistic} and designed its implementation in SPICE~\cite{dowling2020SPICE}.

\begin{figure}[t]
 \includegraphics[width=0.8\columnwidth]{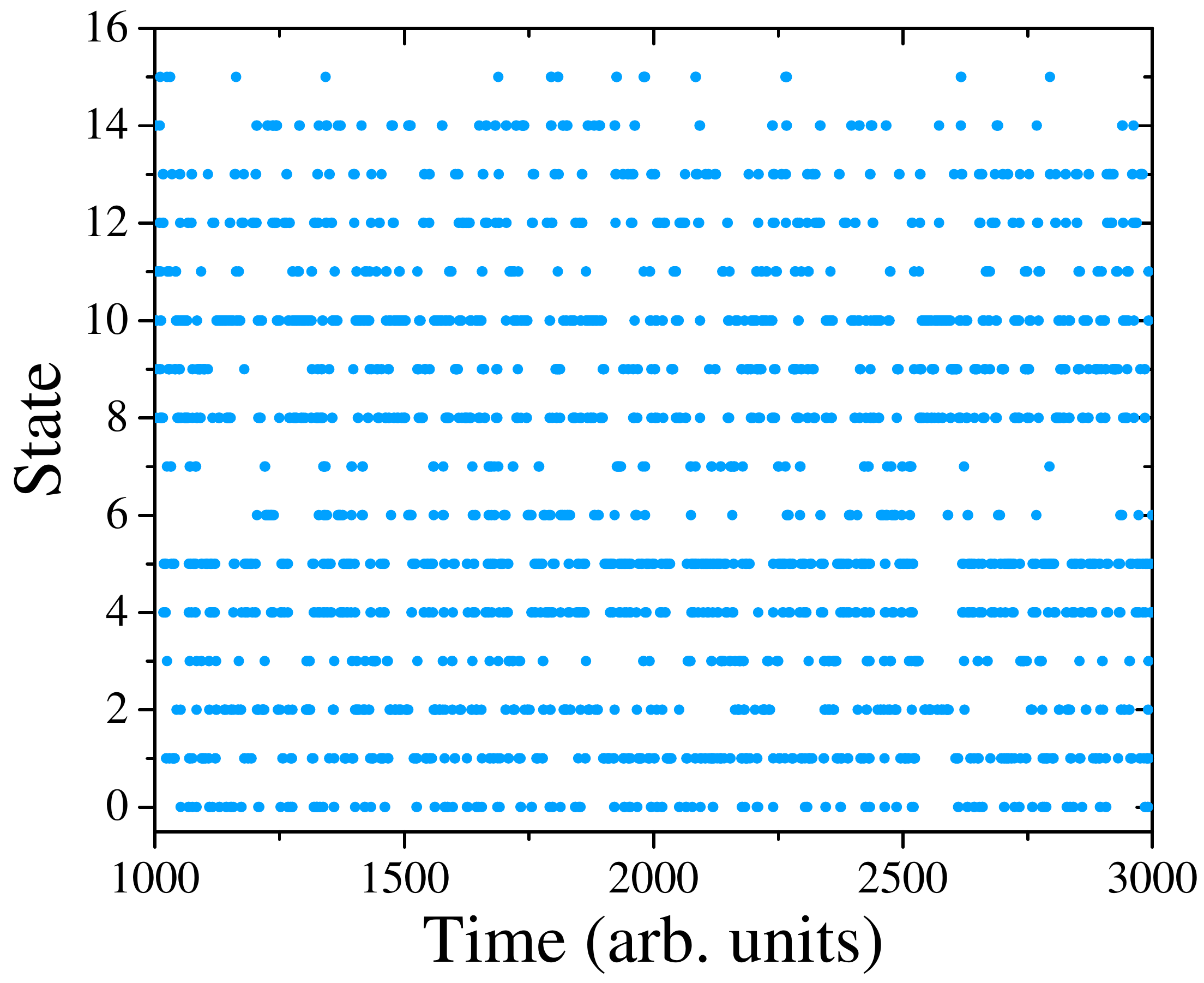} \hspace{2mm}\hspace{2mm}
\caption{ Dynamics of the states in a circuit with $N=4$ memristive spins. The circuit  has $2^4=16$ states that are labeled from 0 to 15. The $0000$ state (all memristors are in $R_{OFF}$) is labeled by 0, $0001$  by 1,  and so on. This plot was obtained using the following set of parameters: $R=r=R_{OFF}=1$~k$\Omega$, $R_{ON}=100$~$\Omega$, $\tau_{01}=3\cdot 10^5$~s, $\tau_{10}=160$~s, $V_{01}=0.05$~V, $V_{10}=0.5$~V, $V_{peak}=1$~V, and $T=2$~s. \label{fig:2}}
\end{figure}

\begin{figure*}[t]
(a)\hspace{2mm} \includegraphics[width=0.8\columnwidth]{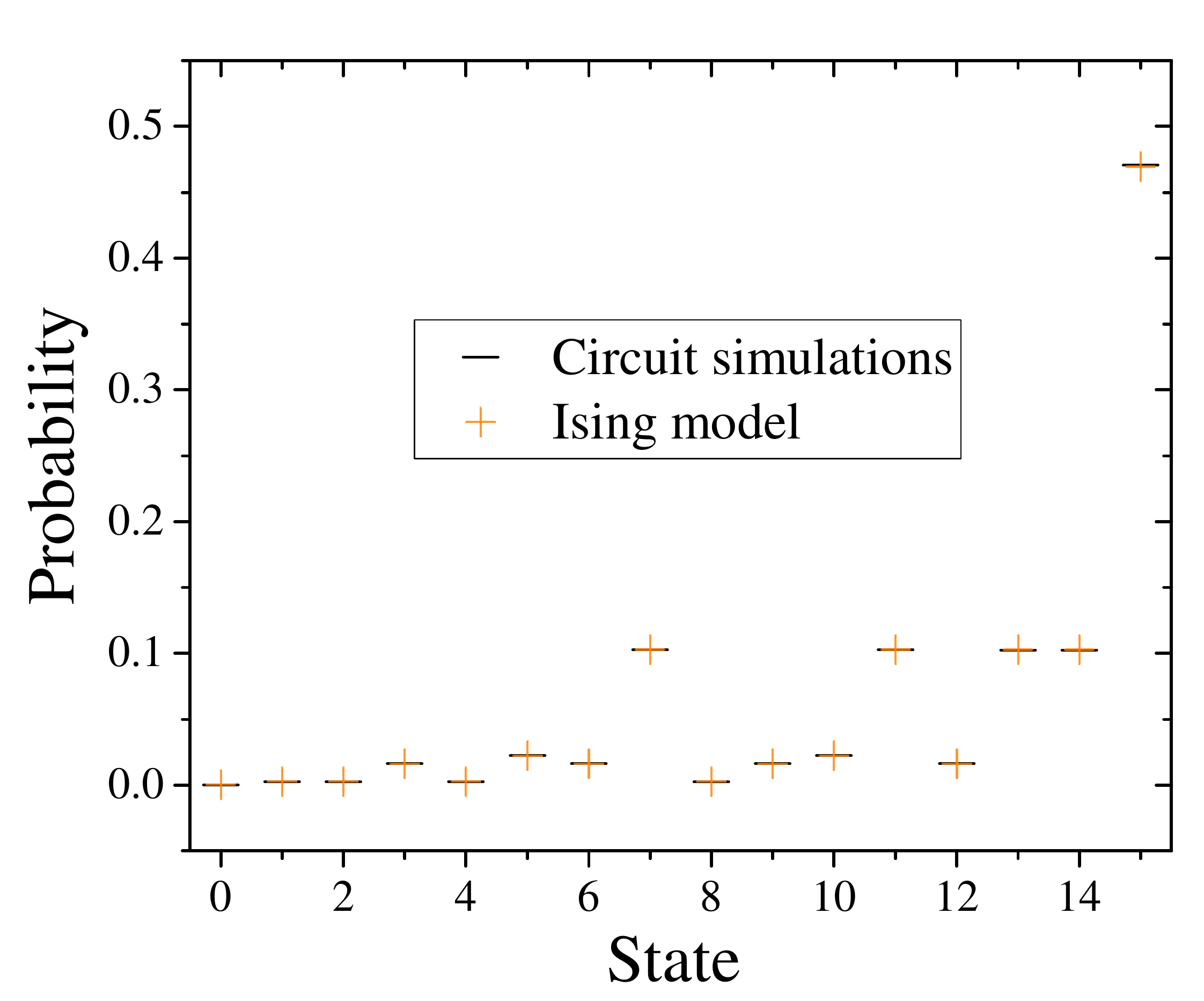}  \hspace{10mm}
(b)\hspace{2mm} \includegraphics[width=0.8\columnwidth]{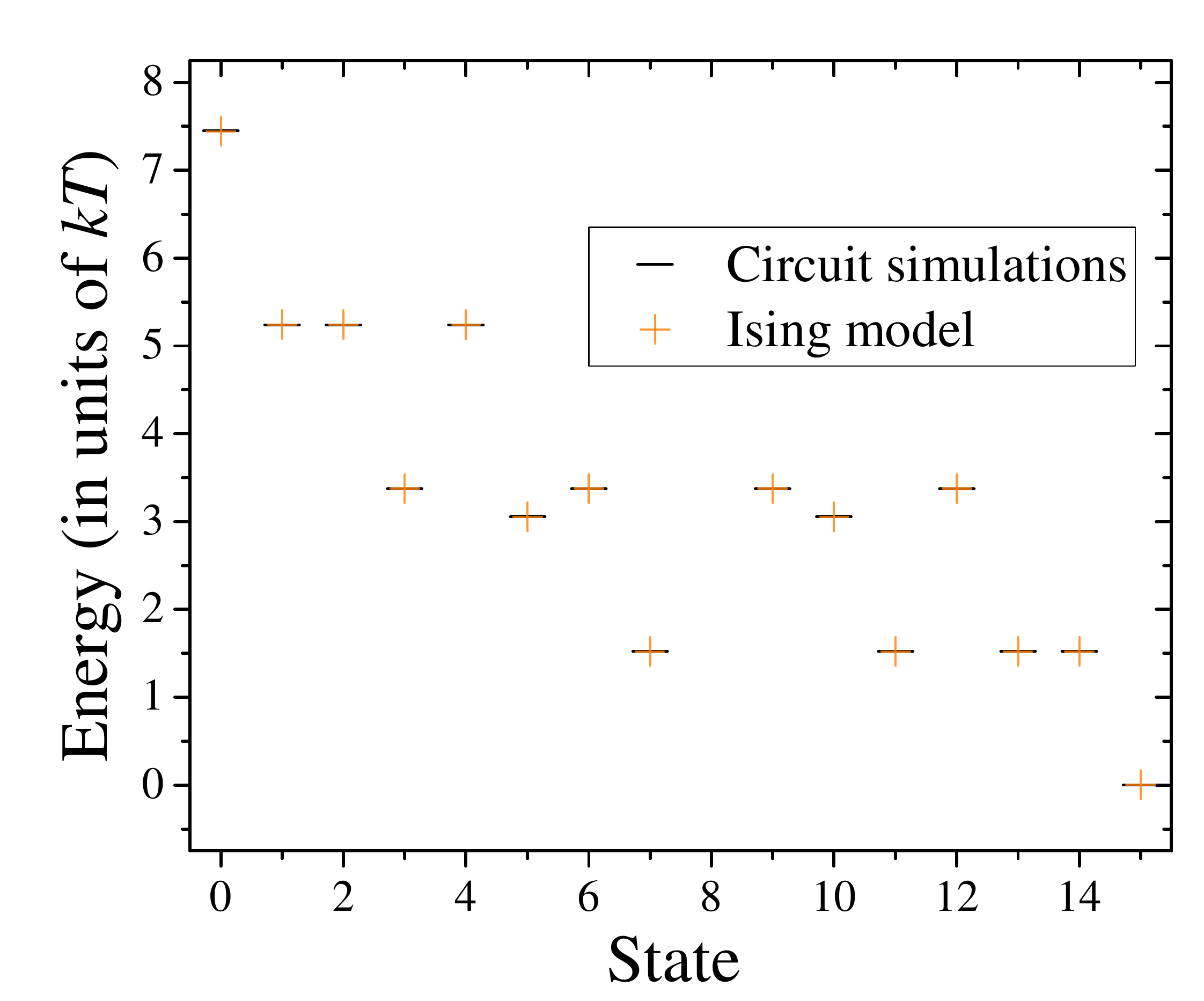} \\
(c)\hspace{2mm} \includegraphics[width=0.8\columnwidth]{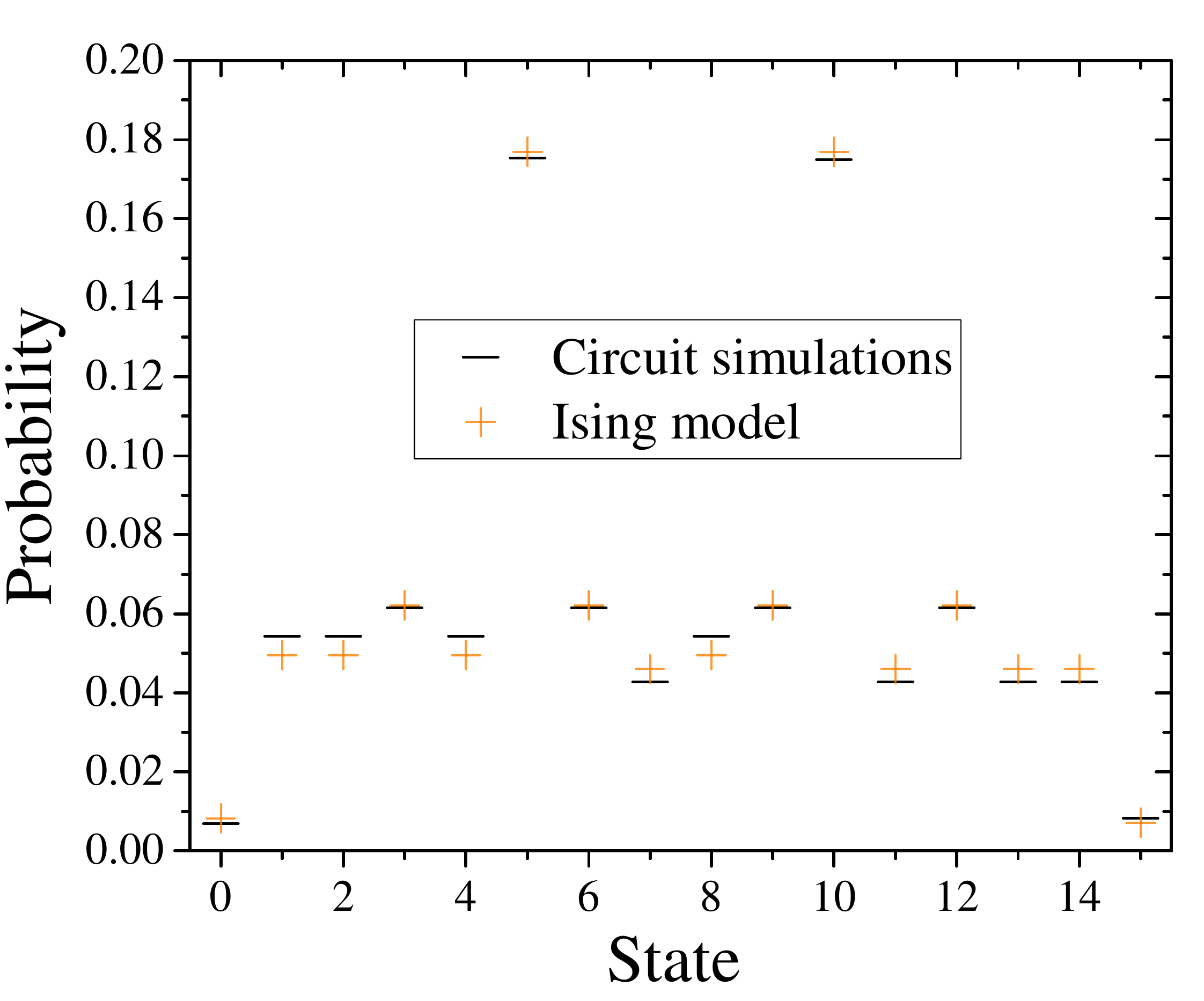}  \hspace{10mm}
(d)\hspace{2mm} \includegraphics[width=0.8\columnwidth]{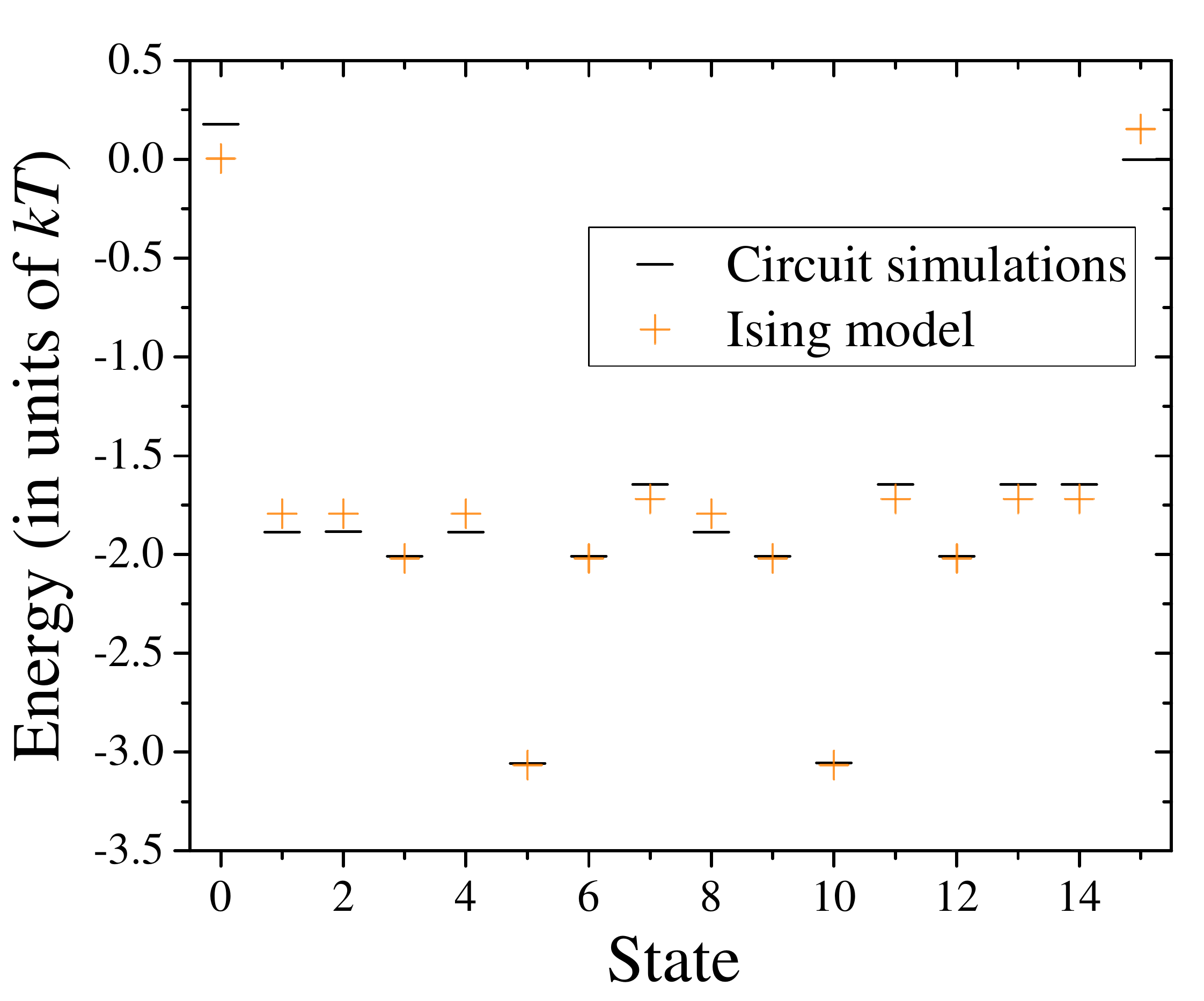}
\caption{Comparison of the probabilities and energies found through a $N=4$ memristor circuit to the values found through the memristor-Ising model for the cases of (a,b) a weaker coupling ($r=10$~k$\Omega$) and (c, d) stronger coupling ($r=1$~k$\Omega$). Other simulation parameters are the same as in Fig.~\ref{fig:2}. For (a), the weaker coupling, coefficient values of $J/kT=-0.0839944$, $h/kT=0.930417 $, and $J_2/kT=-0.0015665$ were found and used. For (b,d), the stronger coupling, coefficient values of $J/kT=-0.195313$, $h/kT=1.35807$, and $J_2/kT=-0.024651$ were found and used. \label{fig:3}}
\end{figure*}

Most of the results presented here are obtained through numerical simulations of the circuit in Fig.~\ref{fig:1}(b) containing $N$ memristive spins.
The set of parameters defining the circuit and the simulations such as the model constants, voltage period, duration, resistances, etc. are first set.
The memristors are then initialized to their starting states (typically all OFF). The voltages across each memristor are calculated for the current time-step through Kirchhoff's laws. The switching time is then generated for each memristor randomly with Eq.~(\ref{eq:distrib}) distribution.
The fastest switching time is extracted and compared to the remaining time in the current voltage pulse. If there is sufficient time remaining in the pulse, that memristor switches states and the the time remaining in the period is decreased. The switching times are generated again. If not, the circuit remains in the same state and the interval of the opposite voltage polarity starts. The simultaneous memristor switchings are not considered as their probability is negligible. 

After a sufficient period of time for the circuit to reach a dynamical steady-state has passed, the memristor configuration will be tracked for each period of the applied voltage. Once the simulation has completed, probabilities for every possible memristor configuration will be found using the distribution of configurations from the simulation. These probabilities can then be utilized to calculate ``energies'' corresponding to the circuit dynamics using Eq.~(\ref{eq:2}).

\section{Results} \label{sec:3}

\subsection{FM and AFM couplings}

Fig.~\ref{fig:2} presents an example of state dynamics in the circuit with 4 memristive spins. One can notice that (on average) the states with antiferromagnetic spin arrangements (such as 5=0101b, 10=1010b, where b denotes base 2 notation) are more occupied compared to the ferromagnetic states (e.g., 6=0110b, 3=0011b, etc.). Consequently, the probability for the antiferromagnetic states is higher and thus one can make the qualitative conclusion that this specific circuit (including the parameters of driving sequence) is described by AFM model ($J<0$).

The Ising model parameters, $J$, $J_2$, and $h$, were found by minimizing the squared difference between the Ising model energies and circuit energies. The latter were obtained based on Eq.~(\ref{eq:2}) that was transformed to $E_i=E_0-kT \ln (p_i/p_0)$. In the Supplemental Material (SI) we provide explicit relations that were used in the calculation of the constants in the Ising Hamiltonian (Eq.~\ref{eq:1}). Fig.~\ref{fig:3} shows the comparison between the probabilities \and energies of the circuit states (found numerically) and ones calculated based on the Ising model. We observed an excellent agreement in the case of a weaker coupling ($r=10$~k$\Omega$), and very good agreement in the stronger coupling case ($r=1$~k$\Omega$).

\begin{figure}
(a)\hspace{2mm} \includegraphics[width=0.8\columnwidth]{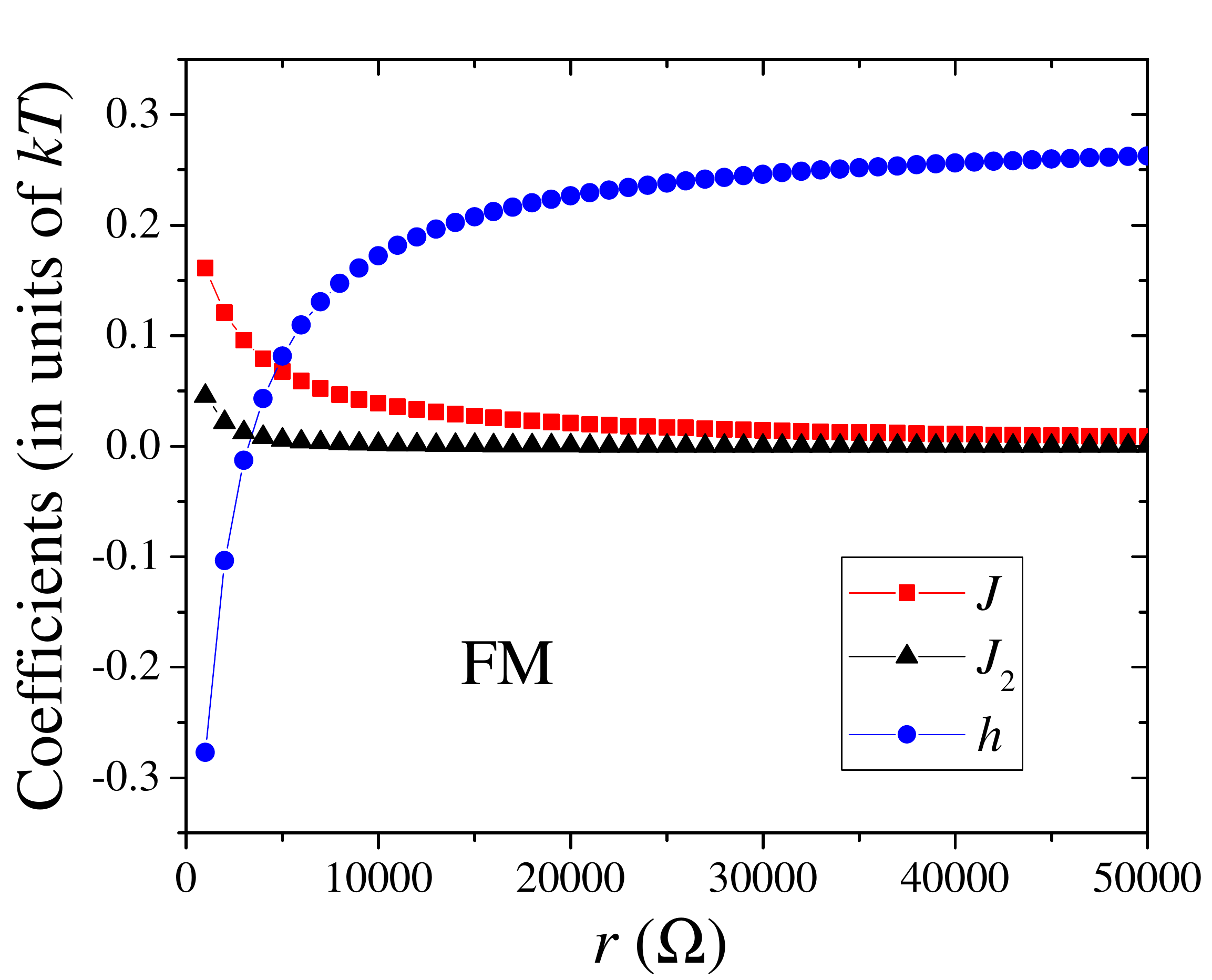} \hspace{2mm}\hspace{2mm}
(b)\hspace{2mm} \includegraphics[width=0.8\columnwidth]{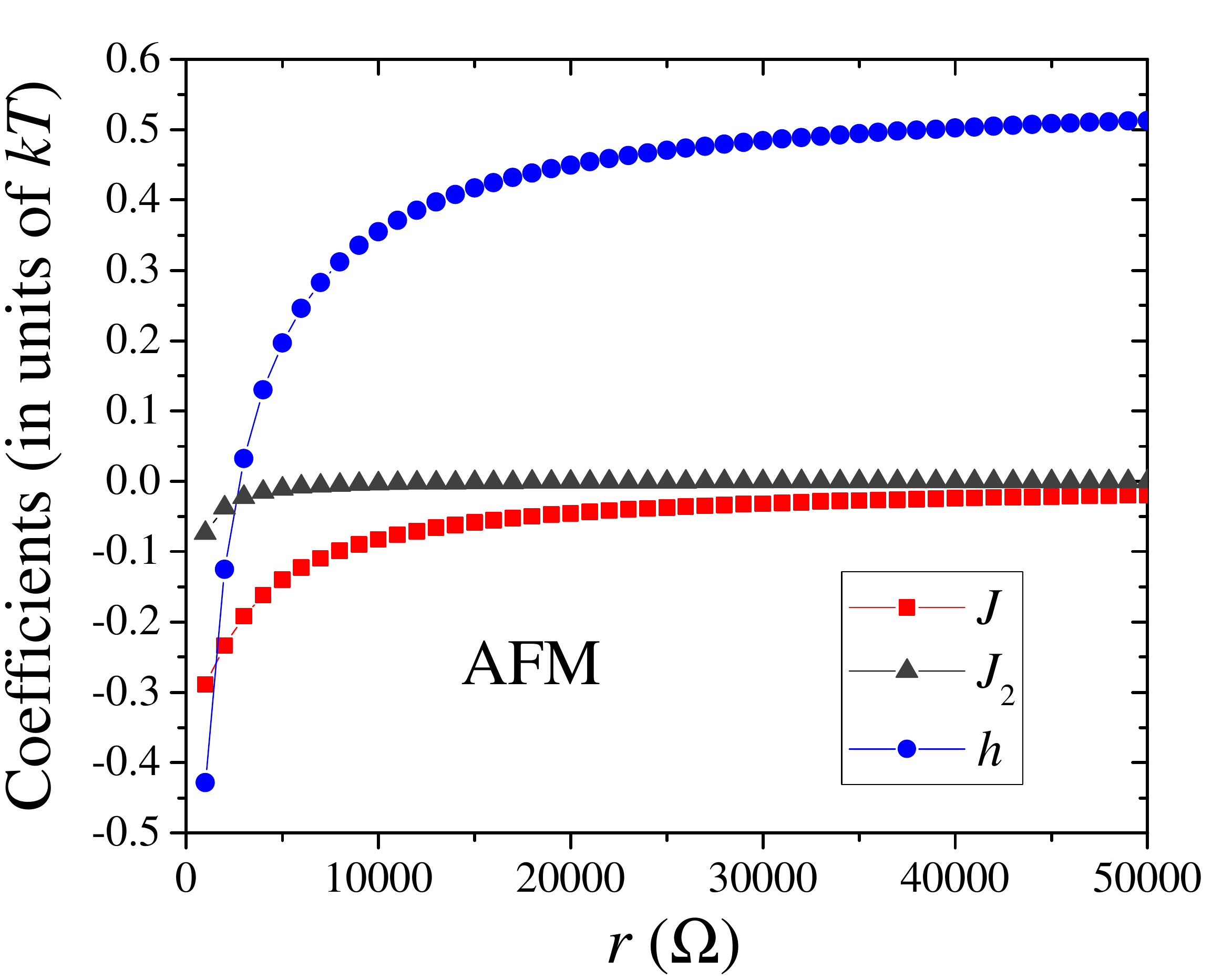}\hspace{2mm}\hspace{2mm}
\caption{Comparison of Ising coefficients using two different sets of model parameters as the coupling resistance $r$ is varied showing (a) FM and (b) AFM interactions of memristive spins. The common parameters are $N=10$,
$R=1$~k$\Omega$, $R_{ON}=500$~$\Omega$, $R_{OFF}=2000$~$\Omega$,  $V_{peak}=1$~V, and $T=2$~s. In (a) we used $\tau_{01}=160$~s, $\tau_{10}=6\cdot 10^4$~s, $V_{01}=0.5$~V, $V_{10}=0.05$~V. In (b) we used $\tau_{01}=10^7$~s, $\tau_{10}=100$~s, $V_{01}=0.05$~V, $V_{10}=0.5$~V.  \label{fig:4}}
\end{figure}

\begin{figure}
\includegraphics[width=0.49\columnwidth]{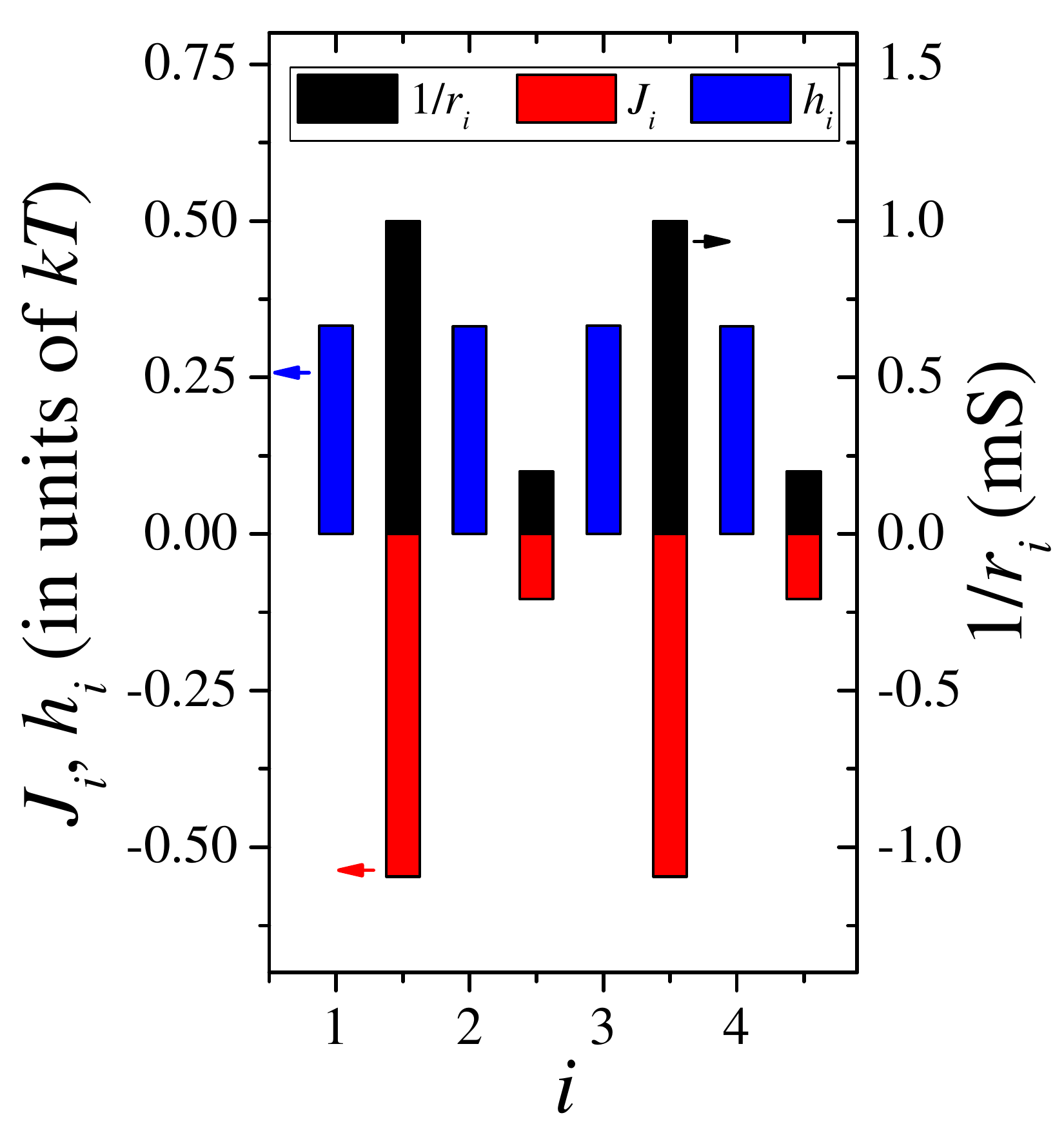}
\includegraphics[width=0.49\columnwidth]{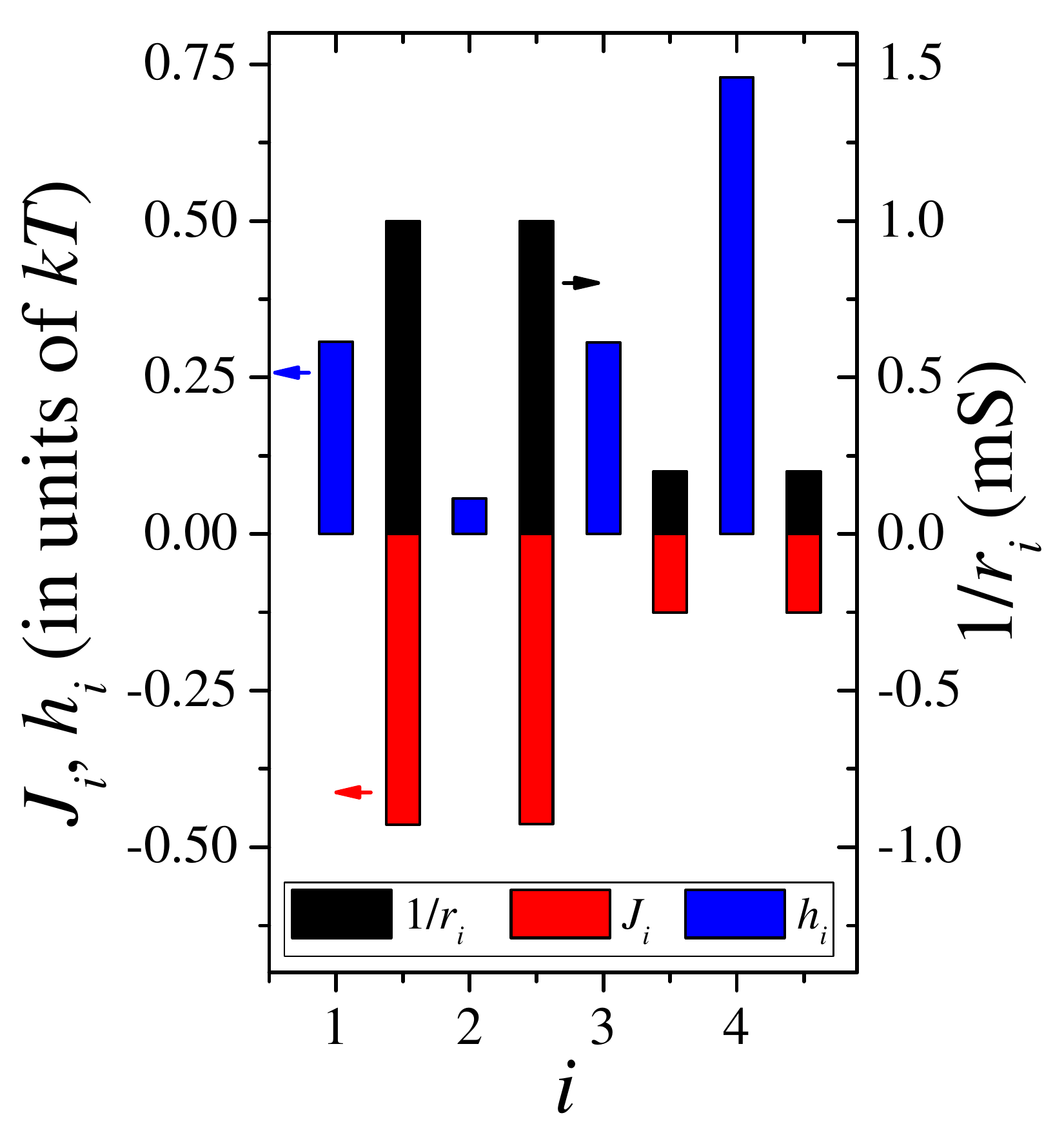} \\
(a) \hspace{4cm} (b) \\
(c) \includegraphics[width=0.49\columnwidth]{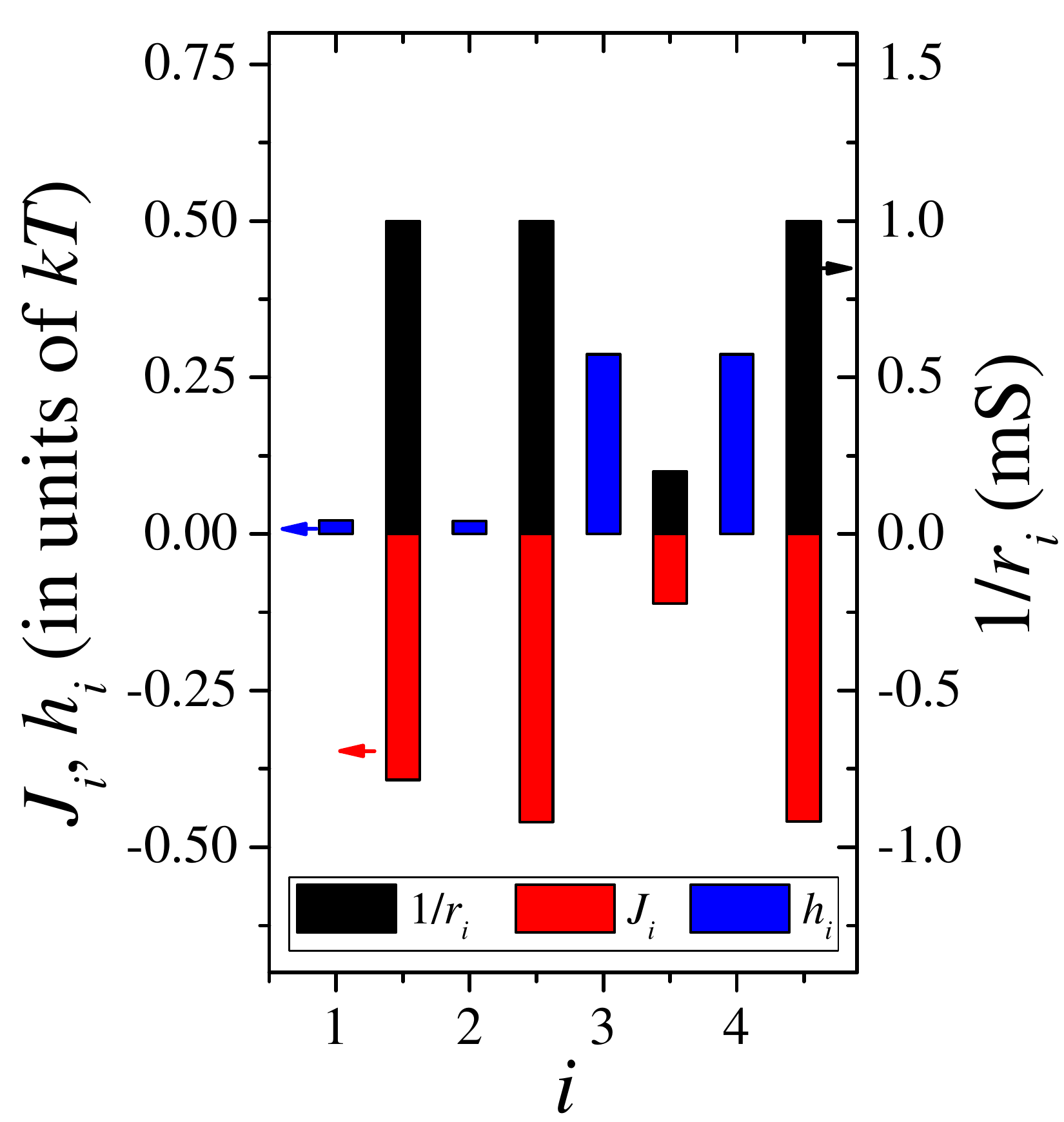}
\caption{Ising model parameters for $N=4$ circuit with distinct coupling resistances $r_i$: (a) $r_1=r_3=1$~k$\Omega$, $r_2=r_4=5$~k$\Omega$; (b)  $r_1=r_2=1$~k$\Omega$, $r_3=r_4=5$~k$\Omega$; (c) $r_1=r_2=r_4=1$~k$\Omega$, $r_3=5$~k$\Omega$. All other simulation parameters are the same as in Fig.~\ref{fig:2}. \label{fig:5}}
\vspace{-0.6cm}
\end{figure}

The main result of this paper can be seen in Fig.~\ref{fig:4}. The figures show how $J$, $J_2$, and $h$ vary in relation to the size of the coupling resistance between memristive spins. Clear ferromagnetic and antiferromagnetic ordering can be seen depending on the choice of circuit parameters. These results can be easily extended to circuits with distinct resistances and memristor parameters. For instance, in Fig.~\ref{fig:5} we present Ising model parameters found for a circuit with distinct coupling resistances $r$. Since a memristive spin has a stronger influence on its neighbors when the coupling resistance is smaller, smaller coupling resistances result in larger Ising coefficient $J_i$ (in Fig.~\ref{fig:5}, $r_i$ and $J_i$ are shifted by $0.5$ to the right to emphasize their role in the spin-spin interaction).

In general, circuits can be set to prefer a specific ordering through the selection of the model parameters $V_{01}$ and $V_{10}$. These parameters, in a sense, set how susceptible a memristor is to the states of its neighbours.
 As memristors switch between resistance states, they induce changes in not only the voltage across themselves, but the voltages across (in principle) all memristors in the chain in accordance with Kirchoff’s circuit laws. The strength of the induced change, or the interaction, weakens as the distance increases from the switching memristor. Through the interplay of the induced changes in the voltages and the chosen set of model parameters, there will be a bias towards a specific type of ordering.

\begin{figure*}
 (a) \includegraphics[width=0.8\columnwidth]{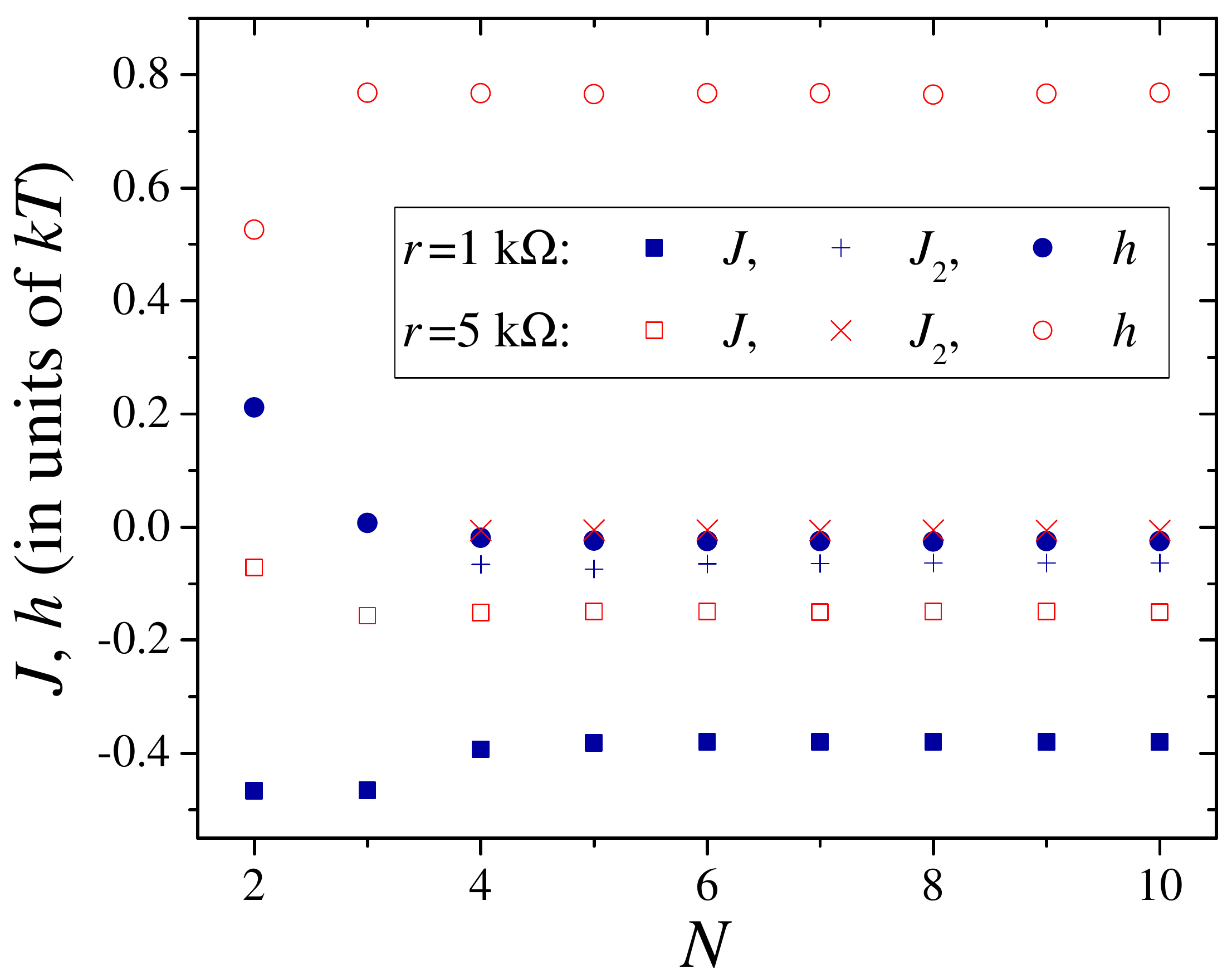} \hspace{2mm}\hspace{2mm}\hspace{2mm}\hspace{2mm}\hspace{2mm}
 (b) \includegraphics[width=0.8\columnwidth]{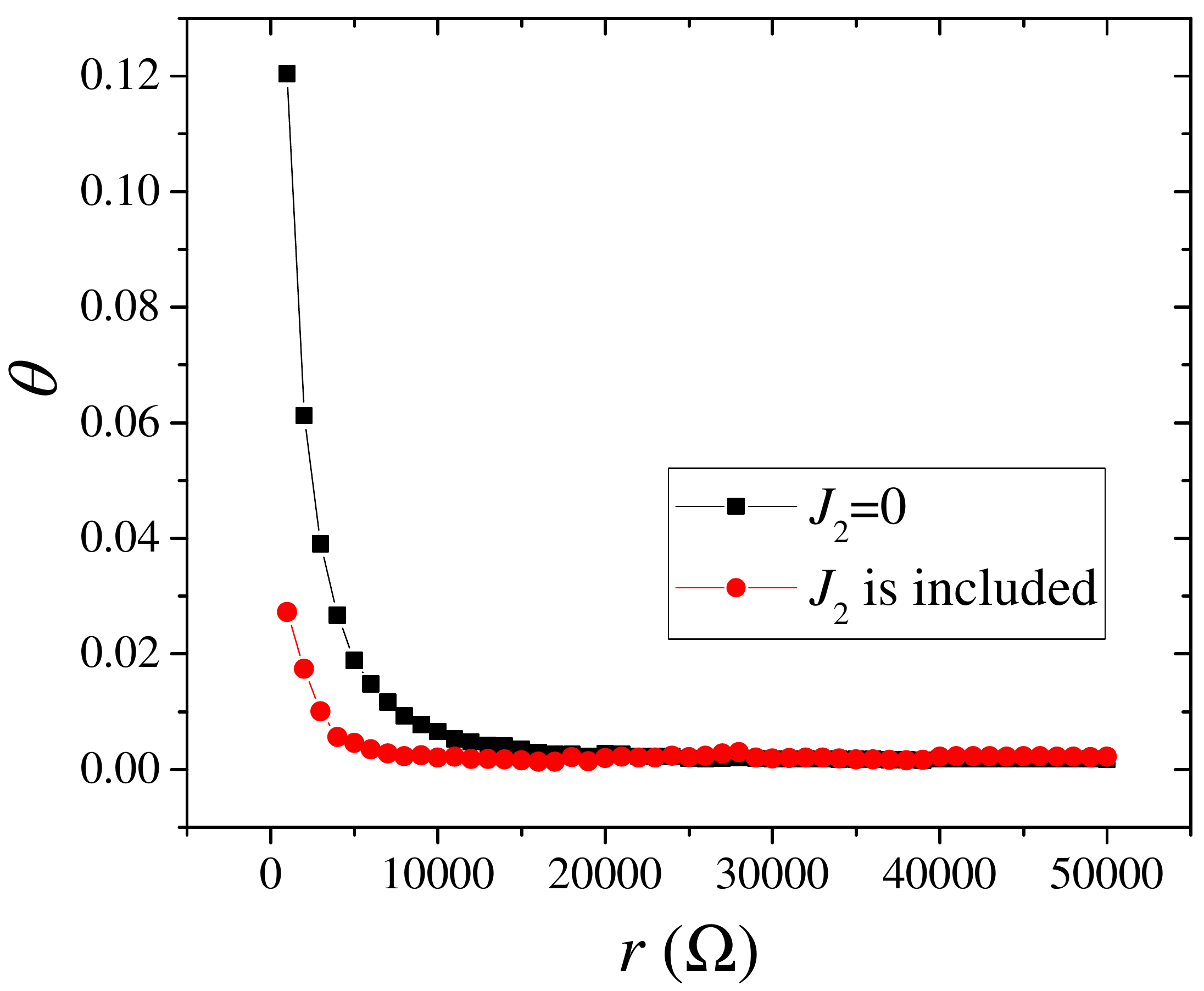}
\caption{(a) Change in the Ising coefficients, $J$, $J_2$ and $h$, as the amount of memristive spins in the circuit is varied from $N=2$ to $N=10$. The simulation parameters are the same as in Fig.~2.
(b) Comparison of the accuracy of the memristor-Ising Hamiltonian with and without $J_2$ using the dot product of the circuit and Ising model probabilities. Performed for the case of AFM calculations in Fig.~\ref{fig:4} with $N=5$.
\label{fig:S2}}
\vspace{-0.6cm}
\end{figure*}

Fig.~\ref{fig:S2}(a) shows the dependency of the Ising coefficients on the amount of memristive spins included within a circuit. Here we can see that at 5 units any major dependency on the amount of units within a circuit disappears. In order to demonstrate the importance of the $J_2$ interaction in  the memristor-Ising Hamiltonian, Fig.~\ref{fig:S2}(b) shows a comparison of approximations with and without $J_2$. It is clear that $J_2$ improves the description only in the stronger coupling case (smaller $r$-s).

\subsection{Comparison with other methods}

As a means of verifying the results seen through numerical simulations, a couple of different methods were employed. For specific cases, meaning specific circuit configurations (generally simplistic), exact solutions can be found for the state probabilities in the master equation~\cite{dowling2020probabilistic}. These results were then compared to the output of the Monte-Carlo simulations to check for agreement. The first method used was exactly solving the master equation analytically through Mathematica. The model parameters were set, the amount of memristors were defined, and all possible memristor voltages for any possible configuration were listed. The switching rates then were constructed for any potential circuit configuration or transition. Using these rates, the master equation was solved exactly for the steady-state~\cite{dowling2022analytic} and the probabilities for each type of memristor configuration were found. The second method used was implementing the master equation in SPICE and using the SPICE environment to find the probabilities for each type of memristor configuration. We have obtained an excellent agreement between the results obtained with different methods (see SI for details).

In order to show agreement, one of the specific configurations considered and directly solved through various means was the circuit in (Fig.~1(b)) containing specifically four memristor-resistor units. This circuit was numerically solved and the master equation was utilized through two applications in order to verify the results obtained by numerical means. In general, the master equation is written as
\begin{equation}
\frac{\textnormal{d}p_{\Theta}(t)}{\textnormal{d}t}=\sum\limits_{m=1}^{N}\left(\gamma_{\Theta_m}^mp_{\Theta_m}(t)-\gamma_\Theta^m p_{\Theta}(t) \right) \;,
\label{eq:10}
\end{equation}
where $p$ is the probability to be in a specific configuration and $\gamma$ is the transition rate between configurations. For the specific case of a circuit with 4 memristor-resistor units the master equations becomes a set of 6 differential equations with forms of (for a fully detailed application of the master equation see~\cite{dowling2020probabilistic})
\begin{equation}
\frac{\textnormal{d}p_{0000}(t)}{\textnormal{d}t}=-4\gamma^1_{0000} p_{0000}+4\gamma^1_{0001} p_{0001} \; . \label{eq:11}
\end{equation}
These differential equations, in conjunction with specific memristor voltages for each possible configuration, can be fully solved in Mathematica and the resulting probabilities for each configuration can be found. A secondary approach is constructing these differential equations in SPICE using a current-controlled voltage source and capacitor pair for each probability along with the full circuit constructed for each memristor configuration~\cite{dowling2020SPICE}. Table S1 shows the probability results for this circuit configuration utilizing the same parameters for all three types on analysis.

\section*{Conclusion}

In conclusion, research into electronic systems that can replicate the statistics of the Ising model or other statistical systems is of increasing interest. In this work, we have demonstrated that
circuits constructed with memristor-resistors units are capable of serving as an analog for the switching behavior exhibited by the Ising model. 
 Our results show an almost perfect match for the energies and probabilities one would expect from the Ising Hamiltonian compared to the numerical simulations performed here. We show the small to negligible gain in accuracy for including interaction terms beyond first neighbor. The results from these simulations were further verified by other forms of analysis. Finally, it is shown that both types of orderings, ferromagnetic and antiferromagnetic, can be realized in these circuits under the appropriate set of model parameters.
 Experimental implementations of the circuits studied would be useful, but they are beyond the scope of this current manuscript.
 This work further adds to the class of electronic circuits that are capable of realizing the behavior of other physical systems.

\bibliography{memcapacitor}

\clearpage
\newpage
\appendix
\section{Supplementary Material: Memristive Ising Circuits}

\setcounter{page}{1}
\setcounter{equation}{0}
\setcounter{figure}{0}
\setcounter{table}{0}
\renewcommand{\theequation}{S.\arabic{equation}}
\renewcommand{\thefigure}{S\arabic{figure}}
\renewcommand{\thetable}{S\arabic{table}}

\subsection{Numerical simulations}

The algorithm used in our simulations can be summarized in the following steps:
   \begin{enumerate}
       \item Set circuit parameters and initialize memristors.
        \item Calculate memristor voltages.
        \item Calculate memristor switching times.
        \item Compare fastest switching time to the remaining voltage period.
        \item Memristor switches if sufficient time is remaining.
        \item If not, the applied voltage changes and next half-period starts.
        \item Track memristor configurations once the circuit reaches equilibrium.
        \item Use distribution of configurations to calculate probabilities and then calculate energies.
    \end{enumerate}

\subsection{Ising model coefficients}
In the Ising model, the energy of a specific lattice configuration can be found from an Ising Hamiltonian, such as Eq.~(1). This Hamiltonian formulation can also be applied to specific stochastic memristor circuits that have units that can be interpreted as spin sites. The connection between the Ising model and dynamics of the chain of memristive spins is explained in the main text.

A general form for these three coefficients can be found through minimizing a cost function defined as
\begin{equation}\label{eq:4}
  F=\sum_{k=0}^{2^N-1}(E_k-E^I_k)^2
\end{equation}
 where $E_k$ is the energy derived from the probability of memristor configuration $k$ and $E^I_k$ is the energy calculated from the Ising Hamiltonian for the same configuration. The cost function can be minimized with respect to each individual coefficient by
\begin{eqnarray}
\frac{\partial F}{\partial J}=
\frac{\partial F}{\partial h}=\frac{\partial F}{\partial J_2}=0\label{eq:5}
\end{eqnarray}
As an example, the form for the coefficient $h$ can be found by the following method
\begin{widetext}
\begin{eqnarray}
\frac{\partial F}{\partial h}&=&\frac{\partial }{\partial h}\sum_{k=0}^{2^N-1}(E_k-E^I_k)^2
=\sum_{k=0}^{2^N-1}\frac{\partial }{\partial h}\left(E_k^2-2E_kE^I_k+(E^I_k)^2\right)\nonumber \\
&=&\sum_{k=0}^{2^N-1}\left[0+2E_k\sum_{i=1}^N\sigma_i+J\sum_{i=1}^N\sigma_i\sigma_{i(\textnormal{mod} N)+1}\sum_{j=1}^N\sigma_j
+J_2\sum_{i=1}^N\sigma_i\sigma_{i(\textnormal{mod} N)+2}\sum_{j=1}^N\sigma_j+2h\sum_{i=1}^N\sigma_i\sum_{j=1}^N\sigma_j\right]
\end{eqnarray}
\end{widetext}
where $\textnormal{mod} N$ is used to account for the periodic boundary conditions in the circuit (memristor $i$~$=$~$N$ has neighbors $i$~$=$~$N-1$ and $i=1$). The terms $\sum\limits_{k=0}^{2^N-1}\sum\limits_{i=1}^N\sigma_i\sigma_{i(\textnormal{mod} N)+1}\sum\limits_{j=1}^N\sigma_j$ and $\sum\limits_{k=0}^{2^N-1}\sum\limits_{i=1}^N\sigma_i\sigma_{i(\textnormal{mod} N)+2}\sum\limits_{j=1}^N\sigma_j$ both equal out to 0. Note that in such expressions the sum over $i$ is performed for the configuration $k$ of memristive spins. Leaving us with the expression
\begin{equation}
h=\frac{-\sum\limits_{k=0}^{2^N-1}E_k\sum\limits_{i=1}^N\sigma_i}{N2^N}\label{eq:6}.
\end{equation}
A similar analysis can be performed for the $J$ coefficient yielding the expression
\begin{equation}
    J=\frac{-\sum\limits_{k=0}^{2^N-1}E_k\sum\limits_{i=1}^N\sigma_i\sigma_{i(\textnormal{mod} N)+1}}{N2^N}\label{eq:7}
\end{equation}
This construction of the memristor-Ising Hamiltonian now allows us to utilize the energies found from numerical simulations to generate Ising coefficients and ultimately derive energies directly from the Hamiltonian.
As in most models, the accuracy of results can be improved by adding additional terms at the cost of increased complexity. In more complicated applications of the Ising model, second neighbor interaction terms and possibly even higher can be included in the Hamiltonian~\cite{Henriques1987Ising}. The same can be done in this memristor-Ising construction. In the circuit, the effect of a memristor changing states is not felt merely by itself and its neighbors, but it permeates throughout the entire circuit. An additional term, $J_2$, can be added to account for the interaction between a memristor unit and its second closest neighboring unit. This additional term will have the form
\begin{equation}
  J_2=\frac{-\sum\limits_{k=0}^{2^N-1}E_k\sum\limits_{i=1}^N\sigma_i\sigma_{i(\textnormal{mod} N)+2}}{N2^N}.\label{eq:9}
\end{equation}

Fig.~6(b) shows a comparison of the model fitting using the Hamiltonian with and without $J_2$. For this purpose, we consider the occupation probabilities as  vectors in $2^N$-dimensional space and calculate the angle between these vectors using their dot product. It is clear that the inclusion of $J_2$ improves the accuracy of the model by an almost negligible amount specifically at higher coupling resistances which suggests implementing even higher order interaction terms is not necessary compared to the added complexity.

\begin{table}[tb]
\caption{Example of probabilities calculated by various means for $N=4$ memristive Ising using the same circuit parameters as in Fig.~2.} 
\centering 
\begin{tabular}{c c c c c c c} 
\hline\hline 
State &  & Mathematica &  & SPICE &  & Numeric\\ [0.5ex] 
\hline 
0000 &  & 0.00584291 &  & 0.005943266 &  & 0.0069041615 \\
0001 &  & 0.05274275 &  & 0.053329550 &  & 0.0543352990 \\
0010 &  & 0.05274275 &  & 0.053329550 &  & 0.0542786650 \\
0011 &  & 0.06166575 &  & 0.061659225 &  & 0.0615103240 \\
0100 &  & 0.05274275 &  & 0.053329550 &  & 0.0543787880  \\
0101 &  & 0.17635500 &  & 0.176309550 &  & 0.1753005600  \\
0110 &  & 0.06166575 &  & 0.061659225 &  & 0.0615064750  \\
0111 &  & 0.04380950 &  & 0.043278525 &  & 0.0428149660 \\
1000 &  & 0.05274275 &  & 0.053329550 &  & 0.0543502950  \\
1001 &  & 0.06166575 &  & 0.061659225 &  & 0.0614562240  \\
1010 &  & 0.17635500 &  & 0.176309550 &  & 0.1749813700  \\
1011 &  & 0.04380950 &  & 0.043278525 &  & 0.0427653890  \\
1100 &  & 0.06166575 &  & 0.061659225 &  & 0.0615269140  \\
1101 &  & 0.04380950 &  & 0.043278525 &  & 0.0428074910  \\
1110 &  & 0.04380950 &  & 0.043278525 &  & 0.0428327070  \\
1111 &  & 0.00857507 &  & 0.009560732 &  & 0.0083684580\\ [1ex] 
\hline 
\end{tabular}
\label{table:Math} 
\end{table}

\subsubsection{Non-uniform chains}

We have also generalized the above results to the case of non-uniform chains. The final expressions are
\begin{equation}
    J_i=\frac{-\sum\limits_{k=0}^{2^N-1}E_k\sigma_i\sigma_{i(\textnormal{mod} N)+1}}{2^N}\label{eq:7aa}
\end{equation}
\begin{equation}
 h_i=\frac{-\sum\limits_{k=0}^{2^N-1}E_k\sigma_i}{2^N}\label{eq:8aa}
\end{equation}
\begin{equation}
  J_{2,i}=\frac{-\sum\limits_{k=0}^{2^N-1}E_k\sigma_i\sigma_{i(\textnormal{mod} N)+2}}{2^N}.\label{eq:9aa}
\end{equation}
The comparison with Eqs.~(\ref{eq:6})-(\ref{eq:9}) shows that Eqs.~(\ref{eq:6})-(\ref{eq:9}) can be considered as Eqs.~(\ref{eq:7aa})-(\ref{eq:9aa}) averaged over all units along the chain.  Note that Eq.~(\ref{eq:7aa}) is valid for $N>2$ and Eq.~(\ref{eq:9aa}) is valid for $N>4$. For $N=2$, $2^N$ should be replaced by $2^{N+1}$ in the denominator in Eq.~(\ref{eq:7aa}). For $N=4$, $2^N$ should be replaced by $2^{N+1}$ in the denominator in Eq.~(\ref{eq:9aa}). Eqs.~(\ref{eq:7aa}) and (\ref{eq:8aa}) were used to obtain Fig.~5.

\end{document}